# SIZE AND COMPOSITION OF CLUSTERS PRODUCED BY SUPERSONIC EXPANSION OF BINARY GAS MIXTURES

O. P. Konotop[1], S. I. Kovalenko, O. G. Danylchenko, and V. N. Samovarov

*B. Verkin Institute for Low Temperature Physics and Engineering*

*of the National Academy of Sciences of Ukraine,*

*47 Lenin Avenue, Kharkiv, 61103, Ukraine*

## ABSTRACT

Size and composition of clusters produced by adiabatic expansion of binary gas mixtures (Ar-Kr, Kr-Xe, and $N_2$-Ar) with various component concentrations are studied by using electron-diffraction technique. The resulting homogeneous and heterogeneous clusters are shown to have compositions substantially different from those of the primary gas mixtures and dependent on cluster size. We have found that the key parameters needed for an analysis of cluster composition are the critical cluster radius and the heavier component concentration in the gas mixture which can be used to establish the regions of existence of homogeneous and heterogeneous clusters. These critical values determine the coefficient of the enrichment of clusters with the heavier component with respect to its concentration in the primary gas mixture. Theoretical relations are obtained which provide a good quantitative description of the experimental results and can be expected to be also valid for finding the composition of binary clusters of other van der Waals systems.

---

[1] Corresponding author: aleksey_konotop@mail.ru

**INTRODUCTION**

Both heterogeneous and homogeneous clusters exhibit a variety of unusual properties that can be characterized as being intrinsic to systems with a limited number of particles and in the bulk materials are either found under extreme conditions or not observed at all. In contrast to homogeneous clusters, properties of heterogeneous clusters can be manipulated by changing not only size, but also component concentration of clusters.

Nowadays, two experimental techniques are used to generate heterogeneous clusters in supersonic jets. According to one of them, a binary gas mixture passes through a nozzle and condenses to produce heated clusters which, while cooling down to a certain temperature, form a cluster beam (the "co-expansion" technique). The other method consists in producing homogeneous clusters and then covering them with a layer of another substance (or layers of different substances) by using atom (molecule) sources (the "pick-up" technique). Clusters obtained by the latter method are characterized by a non-equilibrium component distribution due to the lack of a stage at which hot clusters can be formed, which is needed to allow for a fast reciprocal diffusion of the constituent particles. Below, we only consider clusters generated by the "co-expansion" technique.

A strong impact of the composition of clusters on their physical properties is observed in a number of binary systems, which makes the problem of finding component concentrations in heterogeneous clusters a very important issue in the further development of experimental studies of such systems. Despite the rather low number of the available experimental publications in this field, it can now be said that the component concentration inside binary clusters produced by the "co-expansion" technique can be substantially different from that of the primary gas mixture entering the nozzle. There takes place the so called "enrichment effect" which consists in an increase in the heavier component concentration inside the cluster with respect to its value in the gas mixture. For example, if we vary the Ar mole fraction (mf), $C_{Ar}^{gas}$, in the primary N$_2$-Ar gas mixture from 0.02 to 0.25, the resulting mole fractions inside the cluster will be much higher: from $C_{Ar}^{cl}$ = 0.17 to 0.67, respectively [1]. It follows from another experiment performed on $N_2$-Ar system [2] that for $C_{Ar}^{gas} = 0.1$ mf the Ar concentration in the mixed cluster may amount to 0.35 mf. The enrichment effect has also been observed in large Ar-Kr and Kr-Xe clusters [3] as well as in small Kr-Xe clusters with 30-37 atoms per cluster [4]. In either of the systems, the enrichment coefficient ($\eta_{Kr} = C_{Kr}^{cl}/C_{Kr}^{gas}$ or $\eta_{Xe} = C_{Xe}^{cl}/C_{Xe}^{gas}$, respectively) was much greater than unity, being nonmonotonously dependent on the total pressure and Kr (Xe) concentration in the primary gas mixture [3].

The purpose of the present paper is to perform a more detailed experimental and theoretical study of the enrichment effect in binary clusters. Electron-diffraction measurements were made on Ar-Kr, Kr-Xe, and $N_2$-Ar clusters. Some of the experimental data originate from our previous papers [1,3].

Our experimental results allow us to establish several general relations describing the cluster component composition and enrichment coefficient as functions of cluster size and the heavier component concentration in the primary gas mixture. The studied binary systems are characterized by an almost unlimited solubility in the solid phase and can thus be considered to be good model objects for understanding the enrichment process in van der Waals clusters. In the case of binary systems with a limited solubility of the components, we can expect our results to be valid in the concentration range in which their solid solutions can be formed.

**EXPERIMENTAL TECHNIQUE**

Our experiments were performed on substrate-free mixed clusters produced by condensation of $N_2$-Ar, Ar-Kr, and Kr-Xe gas mixtures in supersonic jets expanding isentropically into a vacuum. The composition, size, and structure of clusters were studied by the electron diffraction method, the experimental setup was described in detail previously [5,6]. The supersonic gas jet expanding into the vacuum chamber of an electron diffraction device was generated by a conical nozzle with the following parameters: the critical cross section diameter was 0.34 mm, the cone opening angle was 8.6°, and the ratio of the areas of the entrance cross section to the critical one was 36.7. To minimize the contribution of the gas component to the diffraction pattern, the mixed cluster-gas jet was collimated at the nozzle exit by a skimmer, a conical diaphragm with an opening of 1.09 mm in diameter.

The diffraction patterns were created by a fraction of the cluster beam at a distance of 100 mm from the nozzle exit, where clusters are already well-formed aggregations concentrated predominantly along the axis line of the jet. The diffraction patterns were registered in a diffraction vector range up to $s = 6$ Å$^{-1}$ ($s = 4\pi\sin\theta/\lambda$, $\theta$ being the Bragg diffraction angle).

Average cluster size could be changed by varying the total pressure $P_0$ and the temperature $T_0$ of the gas mixture at the nozzle entrance: an increase in $P_0$ or a decrease in $T_0$ resulted in greater clusters. The $P_0$ values ranged from 0.05 MPa to 0.6 MPa, $T_0$ was varied from $T_{0\min}$ to 250 K ($T_{0\min}$ = 100, 120, and 190 K for $N_2$ – Ar, Ar-Kr, and Kr – Xe mixtures, respectively).

The average linear characteristic dimension of crystalline clusters, $\delta$ (Å), was extracted from the broadening of the diffraction peaks by using the Scherrer equation and taking into account their additional broadening due to the presence of deformation stacking faults [7,8]. The relative error in calculating $\delta$ did not exceed 10 %. Assuming a spherical form of the clusters, the average number of the constituent particles of a cluster can be calculated as $\langle N \rangle = \frac{\pi\delta^3}{6} \cdot \frac{n}{v}$, where $v$ is the unit cell volume and $n$ is the number of particles in the unit cell. For an fcc unit cell, the equation reads: $\langle N \rangle = \frac{2\pi\delta^3}{3\bar{a}^3}$, where $\bar{a}$ is the measured lattice parameter of mixed clusters. The cluster diameters ranged from 50 to 150 Å ($\langle N \rangle \approx 2\cdot 10^3 - 5\cdot 10^4$ atoms per cluster).

The binary systems studied in the present paper are characterized by an unlimited solubility of the components which leads to the formation of substitutional solid solutions. As was shown in Ref. [9], Vegard's law holds true for our systems in the entire range of their component concentrations. The law relates the mean distance $z$ between the nearest neighbor atoms to the component concentration in the solid solution:

$$z = z_A C_A + z_B(1 - C_A), \tag{1}$$

where $z_A$ and $z_B$ are the shortest distances between atoms of the lattices A and B, while $C_A$ is the concentration of the A component in the solid solution. This means that in order to find the concentration of the heavier components in the cluster, $C_{Ar,Kr,Xe}^{cl}$, one should measure the mean distance $z$ between the nearest neighbor atoms.

**RESULTS AND DISCUSSION**

**The argon-krypton system. Regions of existence of one-component and binary clusters and structural transitions**

It is quite obvious that by keeping unchanged the pressure and the composition of an Ar-Kr mixture at the nozzle entrance but increasing the mixture temperature $T_0$ it is possible to arrive at a situation when only krypton is condensed, since its triple point temperature ($T_{tr}$ = 115.8 K) is higher than that of argon (83.8 K). This immediately poses the question of finding the parameter regions of existence of mixed Ar-Kr and pure Kr clusters. To answer it we did the following experiment. While keeping the gas mixture pressure at $P_0$ = 2 bar for a given krypton concentration $C_{Kr}^{gas}$, we increased the temperature $T_0$ to find the critical value $T_0^{CR}$ above which only pure krypton clusters were formed. Figure 1 shows the measured cluster concentration of krypton $C_{Kr}^{cl}$ versus $T_0$ dependencies for $C_{Kr}^{gas}$ = 0.075 and 0.15 mf, which give the critical temperatures $T_0^{CR}$ = 215 and 177 K, respectively. As the gas mixture temperature was further increased, the size of pure krypton clusters kept decreasing.

All the $T_0^{CR}$ values measured for $C_{Kr}^{gas}$ = 0.025-0.15 mf and $P_0$ = 2 bar and normalized to the Kr triple point temperature are shown in Fig. 2 (solid squares) as function of $C_{Kr}^{gas}$. The hatched area is the region where binary clusters are formed, $T_0 < T_0^{CR}$, above it is observed the region of pure Kr clusters. We can see that starting from $C_{Kr}^{gas} = 0.075$ mf, the experimental data can be fitted by a straight line that in the triple point $T_0/T_{tr} = 1$ yields the Kr critical concentration $C_{CR} = 0.205$ mf above which only pure Kr clusters are formed. The star point in Fig. 2 is $T_0^{CR}/T_{tr} = 2.2$ measured in Ref. [10] for $C_{Kr}^{gas} = 0.08$ mf and a higher pressure value $P_0 = 2.5$ bar. When the gas concentration of krypton is small, the frequency of collisions of Kr atoms drops so strongly that it becomes necessary to significantly increase the gas mixture temperature to have pure clusters of krypton, the boundary between the regions of existence of mixed and pure Kr clusters loses its linearity.

As follows from an analysis of the data of our previous work [3] as well as of the present paper (see below), the critical concentration value $C_{CR}$ does not depend on binary gas mixture pressure $P_0$, all the linear boundaries that separate the regions of mixed and one-component clusters passing through this point. If the heavier component concentrations exceed $C_{CR}$, one-component clusters are formed for all gas mixture pressures. As we will see below, the $C_{CR}$ value for binary clusters is one of the most important parameters that can be used to describe the enrichment effect and find the size of heterogeneous clusters. We believe that the proposed method for finding the regions of existence of homogeneous and heterogeneous clusters can also be used for other binary van der Waals systems which can form solid solutions.

Fig. 2 also shows temperature boundaries of the structural transitions ico – fcc (open circles) and fcc – fcc+hcp (open diamonds) in the Ar-Kr system for $P_0 = 2$ bar. The upper dashed line refers to the transition from a multishell icosahedral (ico) to a crystalline fcc cluster structure. It should be noted that the transition is not sharp, occurring gradually as cluster size grows. It takes place in the cluster size range $\langle N \rangle \approx 2000\text{-}3000$ at/cl. The boundary in Fig. 2 corresponds to the transition onset. The ico and fcc structures are observed for both one-component Kr and binary Ar-Kr clusters. The lower dashed line in Fig. 2 passing through the lower $T_0$ values shows another transition from the fcc to a mixed fcc+hcp structure. The transition is observed in the size range of pure and mixed clusters $\langle N \rangle \approx (2$–$2.5)\,10^4$ at/cl.

**Size of binary Ar-Kr clusters**

To calculate the size of one-component clusters, the following general Hagena relation has long been used [11-13]:

$$\bar{N} = \frac{2\pi}{3}\left(\frac{2R}{a_0}\right)^3 = \gamma \left(\frac{\Gamma^*}{1000}\right)^{\chi},$$

$$\Gamma^* = k_g \left(\frac{0.74 d}{\tan\alpha}\right)^{0.85} \frac{P_0}{T_0^{2.29}} = k_g d_{eq}^{0.85} \frac{P_0}{T_0^{2.29}},\qquad(2)$$

where $a_0$ is the lattice parameter of a homogeneous cluster, $R$ is its radius, $\Gamma^*$ is the empirical Hagena parameter, $P_0$ (mbar) and $T_0$ (K) are the gas parameters at the entrance to the nozzle, $d$ (μm) is the critical diameter of the conical nozzle, $2\alpha$ is the total cone angle, $d_{eq}$ is the equivalent diameter, and $k_g$ is a characteristic constant of the gas ($k_{Xe}= 5500$, $k_{Kr} = 2890$, $k_{Ar} = 1650$, and $k_{N_2} = 528$ [14]). The coefficients $\gamma$ and $\chi$ can be different for various experimental setups with a supersonic nozzle [15]. For our nozzle, the coefficients were experimentally found to be $\gamma = 19.5$ and $\chi = 1.8$.

It follows from the experimental data that the size of heterogeneous Ar-Kr clusters for the Kr gas concentrations $C_{Kr}^{gas} = 0.025$-$0.15$ mf and various $P_0$ and $T_0$ values can be written as follows (see also Ref. [16] for a similar relation):

$$\bar{N} = \frac{2\pi}{3}\left(\frac{2R}{\bar{a}}\right)^3 = 19.5\left[k_{Ar}^{1.8} + \frac{C_{Kr}^{gas}}{C_{CR}}\left(k_{Kr}^{1.8} - k_{Ar}^{1.8}\right)\right]\left\{\frac{d_{eq}^{0.85}}{1000}\frac{P_0}{T_0^{2.29}}\right\}^{1.8\pm0.1},\qquad(3)$$

where $P_0$ is the total gas mixture pressure and $\bar{a}$ is the lattice parameter of a binary cluster. We would like to note that Eq. (3) contains the critical Kr concentration $C_{CR}$ above which only pure Kr clusters are formed. Therefore, at $C_{Kr}^{gas} \geq C_{CR}$ the concentration ratio in the brackets should be set equal to unity to satisfy the condition of the formation of only Kr clusters. If $C_{Kr}^{gas}=0$, Eq. (3) for heterogeneous clusters transforms into Eq. (2) for homogeneous Ar clusters.

**Dependence of krypton concentration on the cluster size**

If we find from our experimental data a relation between the Kr concentration in the cluster and the cluster radius, $C_{Kr}^{cl} = \phi(R)$, then it is possible to calculate the enrichment coefficient $\eta_{Kr}(R) = C_{Kr}^{cl}/C_{Kr}^{gas}$, which can be verified experimentally. On the other hand, using the relation $C_{Kr}^{cl} = \phi(R)$ and Eq. (3), we can arrive at an expression $\eta_{Kr=}\phi(P_0, T_0, C_{Kr}^{gas})$ which relates enrichment coefficient to the parameters of the nozzle and primary gas mixture.

Figure 3a shows the measured $C_{Kr}^{cl}$ values as function of $R$ for the gas mixture Ar(0.9 mf)-Kr(0.1 mf). As clusters grow bigger, the condensation begins with the formation of pure krypton clusters. Starting from the critical radius $R_{CR} = 32$ Å, mixed clusters are formed with krypton concentration $C_{Kr}^{cl}$ decreasing with increasing cluster radius. This behavior of the $C_{Kr}^{cl}(R)$ dependence was found to be typical of the other gas concentrations of krypton studied.

Let us compose an equation allowing us to find the $C_{Kr}^{cl}(R)$ dependence and compare it with the experimental data. Since gas and clusters exchange Ar and Kr atoms through the cluster surface, it is convenient to deal with concentrations per unit area of cluster surface: $\dfrac{C_{Ar}^{gas}}{S} = \dfrac{(1 - C_{Kr}^{gas})}{S}$ (concentration of Ar atoms in gas per unit surface area) and $\dfrac{C_{Kr}^{cl}}{S}$ (concentration of Kr atoms in cluster per unit surface area). The product of the values is proportional to the probability of condensation of an Ar atom accompanied by evaporation of a Kr atom from the cluster. As it has already been shown, at $C_{Kr}^{gas} \geq C_{CR}$ only pure Kr clusters are formed, therefore it is necessary to do the following change: $(1 - C_{Kr}^{gas}) \rightarrow (C_{CR} - C_{Kr}^{gas})$. Then the equation describing the decrease in $C_{Kr}^{cl}$ upon growing cluster volume $V$ reads:

$$\partial C_{Kr}^{cl} = -\beta \frac{(C_{CR} - C_{Kr}^{gas})}{S} \times \frac{C_{Kr}^{cl}}{S} \partial V, \qquad (4)$$

where $\beta$ is a proportionality coefficient. Taking into account the boundary condition $C_{Kr}^{cl} = 1$ mf at $R = R_{CR}$, we arrive at the following solution showing a decrease in Kr concentration in Ar – Kr clusters with growing cluster radius:

$$C_{Kr}^{cl} = \exp\left[\frac{\beta}{4\pi}(C_{CR} - C_{Kr}^{gas}) \times \left(\frac{1}{R} - \frac{1}{R_{CR}}\right)\right] \qquad (5)$$

The solid line in Fig. 3a drawn in accordance with Eq. (5) for $C_{CR} = 0.205$ mf (see Fig. 2), $R_{CR} = 32$ Å, and $\frac{\beta}{4\pi} = 210$ Å is in good agreement with the experimental data. Analysis of the data obtained under various experimental conditions has shown that the $\frac{\beta}{4\pi}$ value depends neither on gas concentrations $C_{Kr}^{gas}$, nor on $P_0$ or $T_0$ and is characteristic of a given binary system.

At the same time, the critical radius $R_{CR}$ does depend on Kr gas concentration. Figure 3b shows the experimental values of $R_{CR}$ for the various Kr concentrations $C_{Kr}^{gas}$. The solid line passing through the experimental points is calculated from the equation below:

$$R_{CR} = \mu (C_{Kr}^{gas})^{0.6}, \qquad (6)$$

where $\mu = 140 \pm 5$ Å. It follows from Fig. 3b that $R_{CR}$ is equal to $R_{tr}$ for the critical concentration of Kr, here $R_{tr}$ is the Kr cluster radius at the triple point given by Eqs. (2) for our nozzle ($R_{tr} = R(T_{tr}, P_{tr}) = 53$ Å). Using Eqs. (2), we can present Eq. (6) as follows:

$$R_{CR}(C_{Kr}^{gas}) = \left(\frac{3\gamma a_0^3}{16\pi}\right)^{\frac{1}{3}} \left(\frac{\Gamma_{tr}^* C_{Kr}^{gas}}{1000 C_{CR}}\right)^{\frac{\chi}{3}}, \qquad (7)$$

where $\Gamma_{tr}^*$ is the Hagena parameter at the Kr triple point, $a_0$ is the lattice parameter for pure Kr, and $\frac{\chi}{3}$ gives the experimentally observed value 0.6. By comparing Eqs. (6) and (7), we can arrive at the following empiric expression for $\mu$:

$$\mu = \left(\frac{3\gamma a_0^3}{16\pi}\right)^{\frac{1}{3}} \left(\frac{\Gamma_{tr}^*}{1000 C_{CR}}\right)^{0.6} = \frac{R_{tr}}{C_{CR}^{0.6}}. \qquad (8)$$

The calculated parameter $\mu = 139$ Å is in good agreement with the experimental value obtained in the present paper ($\mu = 140$ Å). As we are going to see below, Eqs. (7) and (8) are also valid for the systems Kr-Xe and $N_2$-Ar if the Kr parameters are substituted for by those for the respective heavier components (Xe or Ar).

**Enrichment in Ar-Kr clusters**

Figure 4 shows the experimental values of the enrichment coefficient $\eta = C_{Kr}^{cl}/C_{Kr}^{gas}$ as a function of Kr gas concentration $C_{Kr}^{gas}$. To describe quantitatively the results we used the relations discussed above, the principal relation being Eq. (5) with binary cluster radius taken from Eq. (3) and critical radius provided by Eq. (7). The solid curves in Fig. 4 show the dependencies of $\eta$ on $C_{Kr}^{gas}$, which are in good agreement with the experimental results. They were calculated for the following parameters: $\frac{\beta}{4\pi}$ = 210 Å, $\mu$ = 138±2 Å, $C_{CR}$ = 0.205 mf. The lattice parameter of binary clusters used in Eq.(3) was chosen equal to $\bar{a}$ = 5.52 Å, which corresponds to the 0.5 mf concentration of krypton inside the cluster.

We would like to note that all the curves intersect at $C_{CR}$ = 0.205 mf where the enrichment coefficient is close to 5, which actually means that the Kr concentration inside the cluster reaches 1 mf. For gas concentrations $C_{Kr}^{gas} \geq C_{CR}$, only pure krypton clusters are formed in the jet for all $P_0$ and $T_0$.

The $\eta(C_{Kr}^{gas})$ dependencies display pronounced maxima near $C_{Kr}^{gas} \approx 0.05$ mf. The maximum values are pretty large, $\eta$ = 7-9, corresponding to Kr concentrations 0.35-0.45 mf in Ar-Kr clusters. Figure 4 also shows the $\eta(C_{Kr}^{gas})$ dependence calculated for high pressures $P_0 \gg$ 4 bar, when the radius of heterogeneous clusters becomes so large as to exceed greatly $R_{CR}$. In this case, according to Eq. (5), the enrichment coefficient is no more dependent on cluster size. In the vicinity of $C_{Kr}^{gas} \approx 0.05$ mf, the maximum becomes flatter, the coefficient $\eta$ reaches 5 and becomes weakly dependent on $C_{Kr}^{gas}$ for higher concentrations.

Equations (3) and (5) can yield temperature dependencies of various parameters of heterogeneous clusters at a constant pressure. Solid line in Fig. 1 shows the calculated $C_{Kr}^{cl} = \phi(T_0)$ curve for $P_0$ = 2 bar and $C_{Kr}^{gas}$ = 0.075 mf, which fits well the experimental data demonstrating an increase in Kr concentration in clusters with temperature.

Figure 5 shows the size dependencies of the Ar-Kr system enrichment coefficient for two Kr concentrations in the gas mixture, 0.05 and 0.1 mf. Cluster radius was changed by varying the stagnation pressure. It can be seen that at $R > R_{CR}$ the cluster concentration of Ar grows rapidly with radius to flatten out at greater values, $R \approx$ 70-100 Å, where cluster composition becomes weakly dependent on cluster size. Enrichment coefficients obtained for the various $T_0$ and $P_0$ values fall on the same curve $\eta(R)$, whose behavior only depends on the heavier component concentration in the primary gas mixture.

**Enrichment in Kr-Xe clusters**

Let us consider the experimental data and the calculated dependencies $\eta(C_{Xe}^{gas}) = C_{Xe}^{cl}/C_{Xe}^{gas}$ demonstrating the effect of enrichment of Kr-Xe clusters with xenon (see Fig. 6). The $T_0$ value was chosen so that the $T_0/T_{tr}$ ratio is the same for both Kr-Xe and Ar-Kr systems to ensure the same condensation conditions for the two systems, here $T_{tr}$ is the triple point of the heavier component of a binary mixture.

Similarly to Ar-Kr clusters, the $\eta = \phi(C_{Xe}^{gas})$ dependencies display wide maxima at $C_{Xe}^{gas} = 0.06$ mf with $\eta$ amounting to 7-9. The calculated curves fit well the experimental data. To calculate them, we chose the critical concentration of xenon to be $C_{CR} = 0.205$ mf and used the parameter $\frac{\beta}{4\pi} = 225$ Å, which is nearly that for the Ar-Kr system. The solid curves in Fig. 6 were calculated for $\mu = 143 \pm 8$ Å; Eq. (8) gives the value $\mu = 149$ Å. The lattice parameter of the heterogeneous Kr-Xe cluster used in Eq. (3) was chosen to be $a = 5.93$ Å, which, as in the case of Ar-Kr system, corresponds to the 0.5 mf of the heavier component. The lowest curve in Fig. 6 demonstrates the enrichment effect for large clusters, when the condition $R >> R_{CR}$ is satisfied.

**Enrichment in N$_2$-Ar clusters**

To quantitatively describe the enrichment of N$_2$-Ar clusters with Ar, we used the experimental results of Ref. [1], where Ar concentration in the clusters, $C_{Ar}^{cl}$, was measured as function of Ar concentration in the gas mixture, $C_{Ar}^{gas}$. The measurements were performed at $T_0 = 100$ K in the pressure range $P_0$ = 1.5-5 bar. For these parameter values, Ar concentration $C_{Ar}^{cl}$ turns out to be independent of pressure (cluster size), which was demonstrated in our previous measurements [1,3]. Hence in Eq. (5) we should take $R \gg R_{CR}$. It should be noted that, according to the data of Ref. [2], $C_{Ar}^{cl}$ in N$_2$-Ar clusters is independent of $P_0$ at $T_0$ = 300 K and $P_0 \geq 20$ bar. These $T_0$ and $P_0$ values, if shifted along the adiabatic curve for Ar, correspond to the above-mentioned ones from Ref. [1].

Figure 7a shows the measured $C_{Ar}^{cl}$ vs. $C_{Ar}^{gas}$ dependencies. The condensation onset of pure Ar clusters gives the critical value of Ar concentration, $C_{CR}$ = 0.75 mf, needed for further calculations. The solid curve in Fig. 7a shows the dependence of Ar concentration in the cluster on its concentration in the gas mixture calculated from Eq. (5) with the fitting parameter $\frac{\beta}{4\pi} = 23$ Å and $\mu = 63$ Å. The parameter $\mu$ was calculated using Eq. (8), while $R_{CR}$ is given by Eq. (7). The lattice parameter was taken equal to 5.51 Å, which corresponds to the 0.5 mole fractions of the components. The calculated curve fits well the experimental data, including the point taken from Ref. [2] (solid square).

Figure 7b shows the measured and calculated values of the coefficient of the enrichment of N$_2$-Ar clusters with Ar, $\eta = C_{Ar}^{cl}/C_{Ar}^{gas}$, as function of $C_{Ar}^{gas}$. Here too, the lower curve calculated for $R \gg R_{CR}$ (that is when $C_{Ar}^{cl}$ does not depend on cluster radius $R$) is in agreement with the experiment. The upper curve shows the calculated dependence $\eta = \phi(C_{Ar}^{gas})$ when $C_{Ar}^{cl}$ is dependent on $R$ in accordance with Eq. (5). The $\mu$ and $\frac{\beta}{4\pi}$ parameters are the same for both curves. In N$_2$-Ar clusters, the enrichment coefficient has a pronounced maximum near $C_{Ar}^{gas}$ = 0.05 mf, as is the case for Ar-Kr and Kr-Xe clusters.

**CONCLUSION**

The paper deals with an experimental and theoretical study of the composition of binary Ar-Kr, Kr-Xe, and N$_2$-Ar clusters which depends on the various parameters of the primary gas mixture. The main emphasis is given to the size effect on the composition of binary clusters, including their enrichment with the heavier component, whose concentration inside the cluster can be an order of magnitude greater than its value in the gas mixture.

It was found for Ar-Kr and Kr-Xe systems that the enrichment coefficient is strongly dependent on cluster size. As long as cluster radius remains smaller than some critical value $R_{CR}$, only pure clusters of the heavier component are formed in the supersonic jet, but once it exceeds the critical radius ($R > R_{CR}$), mixed clusters enriched with the heavier component are generated. The maximum enrichment is observed at $R \approx R_{CR}$; at greater radii, the enrichment coefficient drops rapidly to become weakly dependent on cluster size. Analysis of the results has shown that a similar size effect should take place in the case of $N_2$-Ar system.

The enrichment coefficient for all three of the systems has been found to pass through a maximum at a certain concentration of the heavier component in the primary gas mixture. If concentration of the heavier component is higher than some critical value $C_{CR}$, only clusters consisting of the heavier component are formed in the jet.

Some theoretical relations have been obtained which describe the composition of clusters and their enrichment as functions of cluster size and component concentrations in the primary gas mixture. We have demonstrated that they are in good agreement with the experimental data and thus can apparently be applied to other binary systems with an unlimited solubility of the components (for example, $CD_4$-Kr, $N_2$-$O_2$, Ar-$O_2$, Kr-$N_2$, $CH_4$-$CD_4$). We expect these relations to be also valid in the case of binary clusters with a limited solubility of the components in the concentration range in which their solid solutions can be formed (for example, $H_2$-Ar, Xe-$CO_2$, Ar-Xe).

**Acknowledgments** The authors are grateful to Dr V. L. Vakula for fruitful discussions and assistance in preparing the paper.

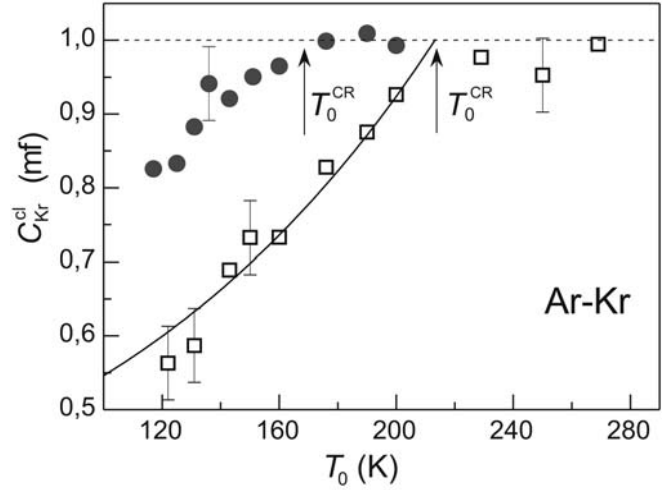

**Fig. 1** Krypton concentration in Ar-Kr clusters as a function of $T_0$ for the Kr concentrations in the primary gas mixture $C_{Kr}^{gas}$ =0.075 mf (open squares) and 0.15 mf (solid circles). Stagnation pressure is $P_0 = 2$ bar. The arrows show the critical temperatures, $T_0 = T_0^{CR}$, above which only virtually pure clusters of krypton are formed. The solid curve is drawn in accordance with Eq. (5).

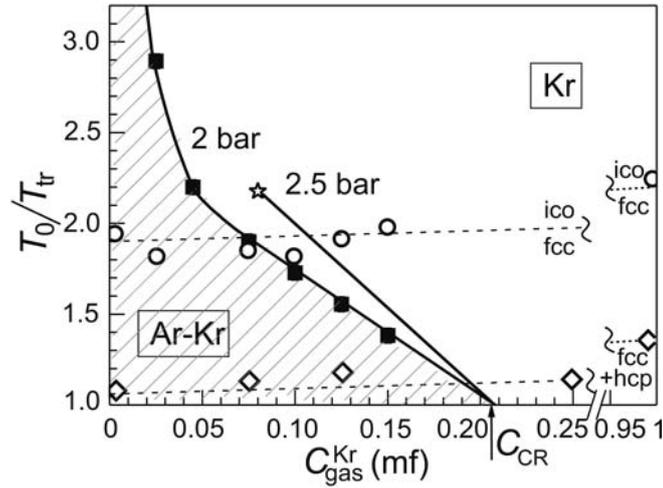

**Fig. 2** Critical temperature $T_0^{CR}$ of the formation of pure Kr clusters normalized to the Kr triple point $T_{tr}$ as a function of Kr concentration in the primary gas mixture for $P_0 = 2$ bar (solid squares). The asterisk shows the data of Ref. [10] for $P_0 = 2.5$ bar. The cross-hatched area corresponds to the region in which binary Ar-Kr clusters are formed. The temperature boundaries of the ico – fcc and fcc – fcc+hcp transitions in the Ar-Kr system for $P_0 = 2$ bar are shown as open circles and open diamonds, respectively.

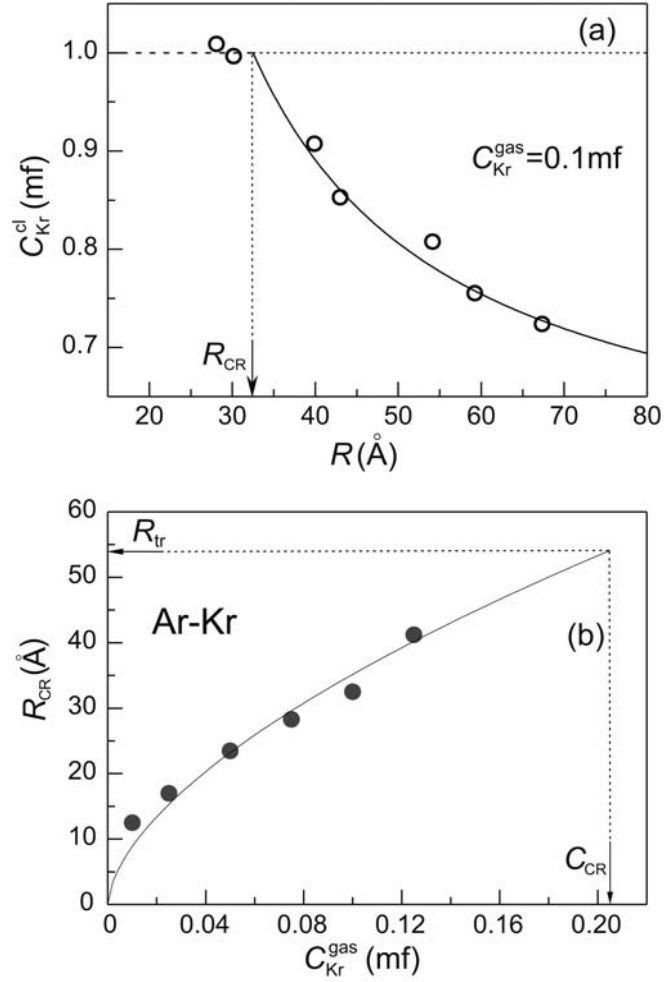

**Fig. 3** (a) Krypton concentration in the mixed Ar-Kr clusters ($C_{Kr}^{gas} = 0.1$ mf) versus cluster radius for $T_0 = 175$ K. The solid curve is calculated from Eq. (5). (b) Critical radius $R_{CR}$ of Ar-Kr clusters versus concentration $C_{Kr}^{gas}$. The solid curve represents the relation $R_{CR} = 140(C_{Kr}^{gas})^{0.6}$.

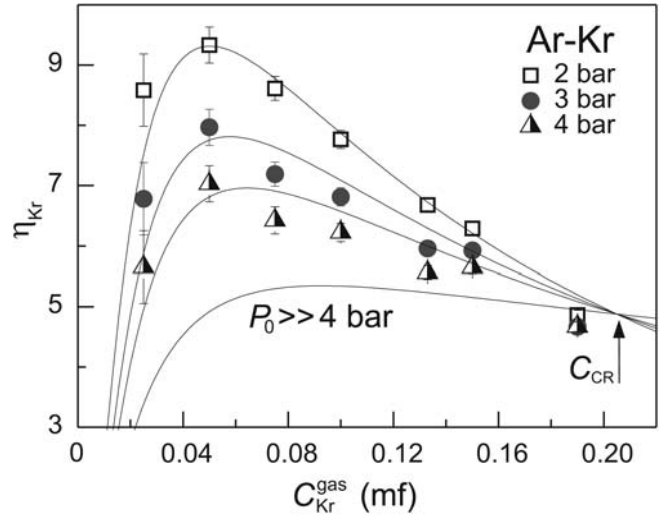

**Fig. 4** Coefficient of the enrichment of mixed Ar-Kr clusters with Kr, $\eta_{Kr} = C_{Kr}^{cl}/C_{Kr}^{gas}$, as a function of $C_{Kr}^{gas}$ for $T_0 = 140$ K and various $P_0$ values. Solid curves are calculated from Eqs. (3) and (5).

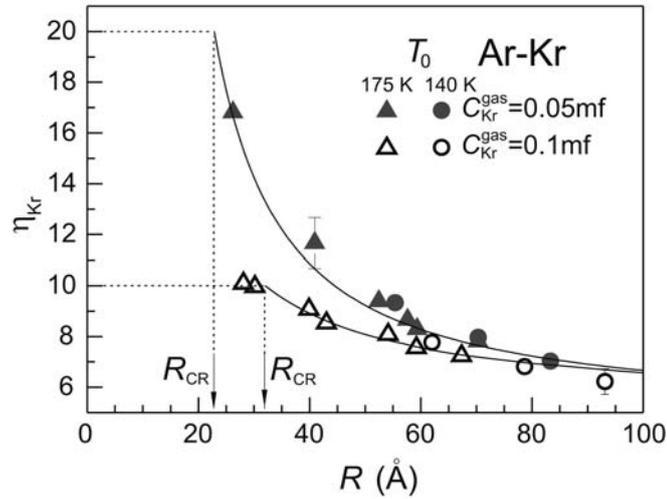

**Fig. 5** Coefficient of the enrichment of mixed Ar-Kr clusters with Kr, $\eta_{Kr}$, as a function of cluster radius for $C_{Kr}^{gas} = 0.05$ (solid circles, $T_0 = 140$ K; solid triangles, $T_0 = 175$ K) and 0.1 mf (open circles, $T_0 = 140$ K; open triangles, $T_0 = 175$ K). Solid curves are calculated from Eq. (5).

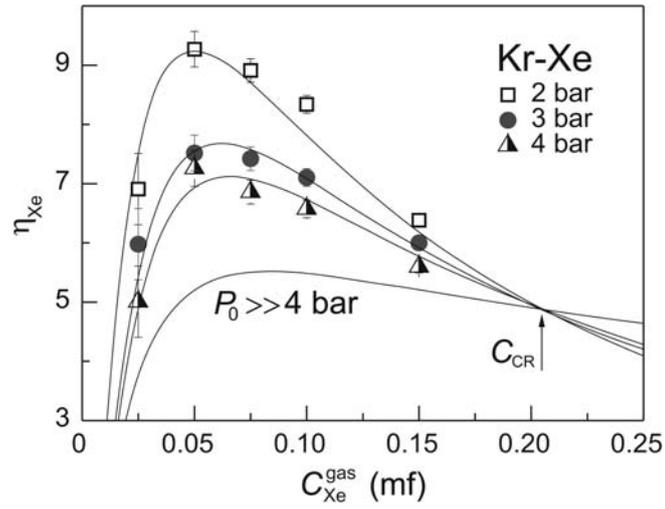

**Fig. 6** Coefficient of the enrichment of mixed Kr-Xe clusters with Xe, $\eta_{Xe}$, as a function of $C_{Xe}^{gas}$ for $T_0 = 193$ K and various $P_0$ values. Solid curves are calculated from Eqs. (3) and (5).

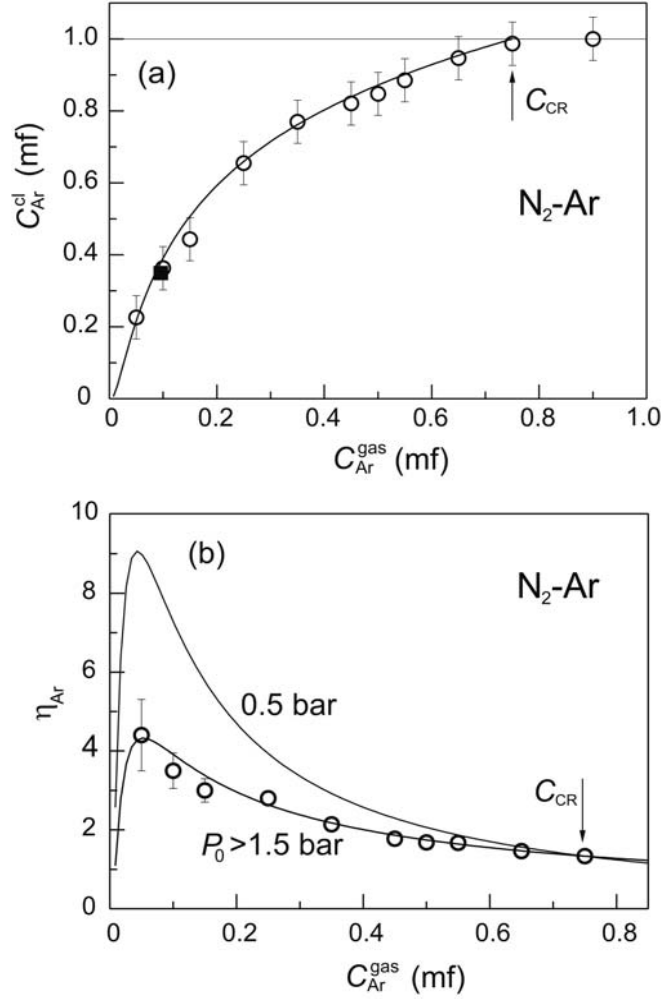

**Fig. 7** (a) Argon concentration in the mixed $N_2$-Ar clusters as a function of $C_{Ar}^{gas}$ for $T_0 = 100$ K and $P_0 > 1.5$ bar (data shown as open circles are taken from Ref. [1], the point shown as solid square is taken from Ref. [2]). (b) Coefficient of the enrichment of mixed $N_2$-Ar clusters with Ar, $\eta_{Ar}$, as a function of $C_{Ar}^{gas}$ for $T_0 = 100$ K. Solid curves are calculated from Eqs. (3) and (5).